\documentclass[%
 aip,
 amsmath,amssymb,
 reprint,%
]{revtex4-1}

\usepackage{graphicx}
\usepackage{dcolumn}
\usepackage{bm}

\usepackage[utf8]{inputenc}
\usepackage[T1]{fontenc}
\usepackage{mathptmx}
\usepackage{textcomp}

\begin{document}

\preprint{AIP/123-QED}

\title[McFadden et. al.]{Epitaxial Al/GaAs/Al tri-layers fabricated using a novel wafer-bonding technique}

\author{Anthony P. McFadden}
\author{Aranya Goswami}%

\affiliation{ 
Department of Electrical and Computer Engineering, University of California, Santa Barbara, CA 93106
}%

\author{Michael Seas}
\affiliation{%
Materials Department, University of California, Santa Barbara, CA 93106
}%

\author{Corey Rae H. McRae}
\affiliation{%
Physics Department, University of Colorado, Boulder, CO 80309
}%
\affiliation{%
National Institute of Standards and Technology, Boulder, CO 80305
}%

\author{Ruichen Zhao}
\affiliation{%
Physics Department, University of Colorado, Boulder, CO 80309
}%
\affiliation{%
National Institute of Standards and Technology, Boulder, CO 80305
}%

\author{David P. Pappas}
\affiliation{%
Physics Department, University of Colorado, Boulder, CO 80309
}%
\affiliation{%
National Institute of Standards and Technology, Boulder, CO 80305
}%

\author{Christopher J. Palmstr\o m}
\affiliation{ 
Department of Electrical and Computer Engineering, University of California, Santa Barbara, CA 93106
}%
\affiliation{%
Materials Department, University of California, Santa Barbara, CA 93106
}%

\date{\today}

\begin{abstract}
Epitaxial Al/GaAs/Al structures having controlled thickness of high-quality GaAs and pristine interfaces have been fabricated using a wafer-bonding technique. III-V semiconductor/Al structures are grown by molecular beam epitaxy on III-V semiconductor substrates and bonded to silicon and sapphire. Selective etching is used to remove the III-V substrate followed by surface cleaning and superconductor regrowth, resulting in epitaxial Al/GaAs/Al tri-layers on sapphire or silicon substrates. Structures are characterized with reflection high energy electron diffraction, atomic force microscopy, X-ray photoelectron spectroscopy, transmission electron microscopy, and X-ray diffraction. Applications of these structures to the field of quantum information processing is discussed.
\end{abstract}

\maketitle

\section{\label{sec:level1}Introduction}

Superconductor heterostructures are playing an important role in the rapidly developing field of quantum information processing. Superconducting qubits utilize superconducting circuit elements and Josephson junctions (JJs) which are made using nanofabrication techniques to store information\cite{Krantz2019,Wendin2017}. Transmon qubits which consist of capacitively shunted JJs are made using a variety of superconductors typically deposited on low-loss substrates such as silicon and sapphire. Voltage tunable ‘gatemon’ qubit structures have been demonstrated using Josephson junctions consisting of superconducting aluminum contacts to high mobility InAs nanowires\cite{Larsen2015} and 2D electron gasses\cite{Casparis2018}. In reports exploring topological superconductivity motivated by topological quantum computation, high mobility III-V materials with spin-orbit interaction are integrated with superconductors (most commonly aluminum) in fabricated mesoscopic structures consisting of semiconducting channels, superconducting contacts, and gates\cite{Mourik2012,Shabani2016,Karzig2017}.

In all of the above-mentioned applications, the observed physics and device properties have been shown to be quite sensitive to the superconductor-semiconductor interface. Mechanisms of loss and decoherence in superconducting qubits have been studied extensively\cite{Wisbey2010,Vissers2012,Wang2015,Woods2019}, and it has been shown that material imperfections at superconductor interfaces can be the dominant source of loss\cite{Wang2015}. As a result, large planar capacitors are commonly used in transmon circuits to dilute surface and interface loss which increase coherence at the cost of scalability\cite{Gambetta2017}. In mesoscopic superconductor-semiconductor structures employed to study topological superconductivity, a so called ‘hard’ induced superconducting gap has been shown to be a requirement for topological protection of the proposed topological qubits\cite{Cheng2012,Rainis2012}. These hard induced gaps have only been observed when the superconductor/semiconductor interface is clean and free of oxides\cite{Chang2015,Kjaergaard2016}.

Epitaxial JJ structures are of particular interest in the field of superconducting qubits where the Al/AlO$_x$/Al JJ elements used in state-of-the-art transmons contain a high density of structural and chemical inhomogeneities\cite{Fritz2019,Zeng2015} which contribute to loss and decoherence. High quality   superconductor/dielectric/superconductor tri-layers could replace the JJ and capacitor elements currently used in transmon designs, improving scalability and coherence. Transmons could even be made from a single superconductor/dielectric/superconductor tri-layer having the appropriate dimensions with a single structure serving as both the capacitor and the Josephson junction\cite{Tahan2019}, if the loss associated with the dielectric and interfaces could be made low enough, a challenge which has not yet been overcome. Conventional semiconductors could be an excellent candidate for this application, owing to their well understood and tunable properties including bandgap energy and lattice constant.

Growth of many superconductors on semiconductors including Al-on-GaAs\cite{Cho1978,Petroff1981,Ludeke1981,Pilkington1999} and Al-on-Si\cite{Yamada1984,McSkimming2017} is well established. However, growth of high-quality single crystal semiconductors such as Si and III-V’s on conventional elemental superconductors such as Al is likely not possible due to concerns such as symmetry mismatch and reactions and roughening that occur at the high growth temperature required for the semiconductor. Yan et al.\cite{Yan2018} succeeded in growth of high quality nitride-based semiconductor structures on superconducting NbN, but the authors used thick AlN/GaN buffer layers grown on the NbN and did not address the problem of moving to thin semiconductor layers required for tunneling junctions. Because of the challenges of materials integration, single-crystal semiconductor based tunnel junctions have not been fabricated by direct growth of semiconductors on superconductors, though several other techniques have been used to create tunneling junctions using semiconductors.

JJ structures using amorphous Si\cite{Smith1983,Olaya2009} and Ge\cite{Gibson1985} have been successfully fabricated and studied; however the semiconductors are amorphous which limits their usefulness for superconducting qubit technology due to loss associated with the structural and bonding disorder. Van Huffelen et al.\cite{Huffelen1993} have studied transport in Nb/Si/Nb JJ's with single-crystal degenerately doped p-Si layers. These structures were fabricated using a backside selective etch to form a thin membrane which was subsequently coated with Nb on both sides. Because of the fabrication technique, the thickness of the Si layers could not be precisely controlled in contrast to direct growth techniques and high Boron doping which is required to obtain etch selectivity.

Magnetic tunnel junctions (MTJs) having single-crystal GaAs tunneling barriers were fabricated by Kreozer et al.\cite{Kreuzer2001,Kreuzer2002}. In these studies, an adhesive wafer-bonding technique was used to obtain pinhole free tunnel junctions with single crystal GaAs barriers and poly-crystalline Fe electrodes. Performance of the Fe/GaAs/Fe MTJs was limited by the imperfect magnetic structure at the Fe/GaAs interfaces due to reactions that occur between Fe and GaAs\cite{Kreuzer2002,Zenger2004}.

In this work, a novel wafer bonding and aluminum regrowth process is presented for fabrication of epitaxial Al/GaAs/Al tri-layer structures. These structures have been made with arbitrary GaAs thickness and are shown to be structurally homogeneous with clean, atomically sharp interfaces. Though GaAs was the only semiconductor used in this study owing to the ease of materials growth and established selective etches, it is expected that much of the presented process may be extended to other semiconductors including other III-V’s and Si as well as semiconductor heterostructures.

\section{\label{sec:level1}Process Overview}

\begin{figure*}
\includegraphics{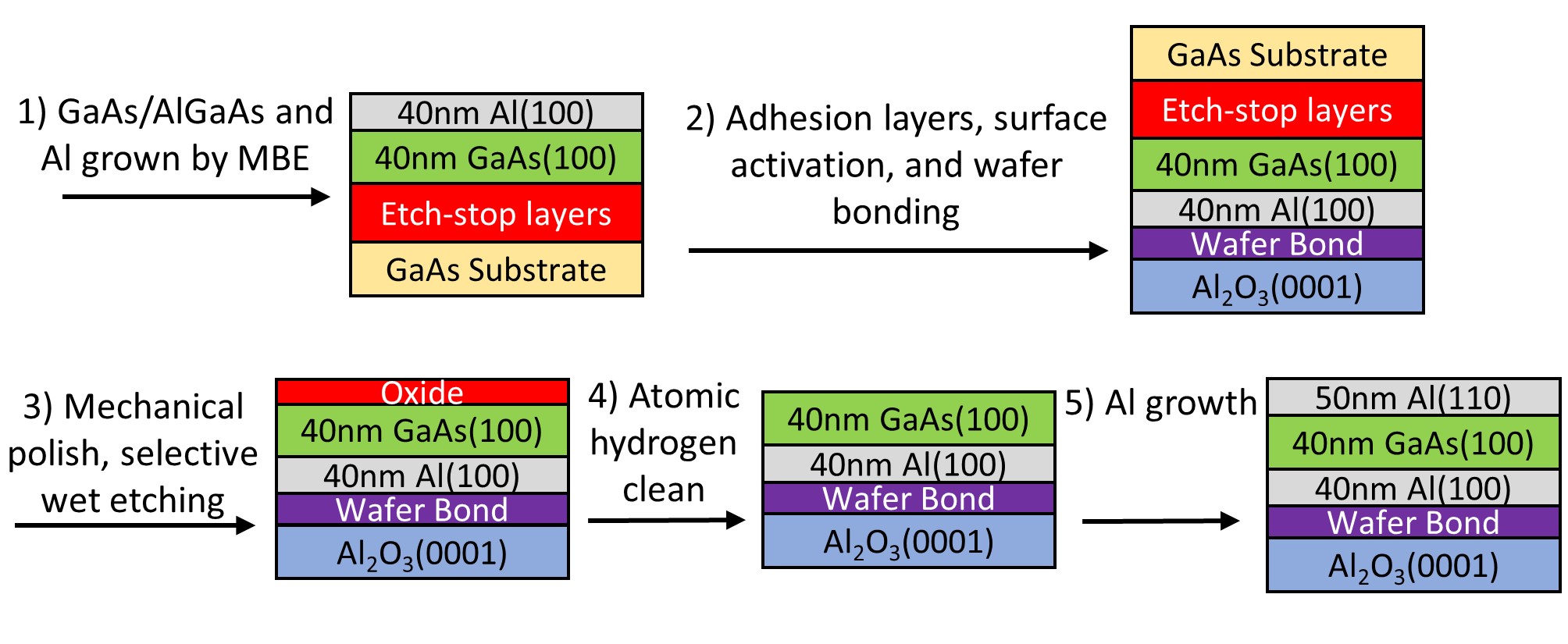}
\caption{\label{Overview}Overview of the Al/GaAs/Al tri-layer process flow.}
\end{figure*}

The novel process used to fabricate epitaxial Al/GaAs/Al tri-layers is outlined in Fig.~\ref{Overview}. 
GaAs/AlGaAs structures grown by MBE are capped with epitaxial Al and removed from vacuum. A wafer bonding process is used to bond the III-V/Al stack to either Si(100) or Al$_2$O$_3$(0001) after which the GaAs(100) substrate is removed using selective wet etching. Following removal of the final AlGaAs protective layer, samples are loaded back into ultrahigh vacuum, the surface oxide is removed using an atomic hydrogen cleaning procedure, and epi-Al is regrown resulting in the complete Al/GaAs/Al tri-layer structure.

\section{\label{sec:level1}Materials Growth}

A Veeco Gen3 III-V MBE system was used for both arsenide and aluminum (superconductor) growth. Semi-insulating, epi-ready 2-inch diameter GaAs(001) substrates were baked at 200\textdegree C in a high vacuum loadlock before outgassing at 350\textdegree C in an ultrahigh vacuum (UHV) preparation chamber. Outgassed substrates were loaded in to the MBE chamber followed by native oxide desorption at 600\textdegree C in an As$_2$ overpressure. GaAs/AlGaAs ‘double-etch-stop’ structures were grown at 580\textdegree C having the layer structure: GaAs(40nm)/Al$_{0.75}$GaAs(50nm)/GaAs(250nm)/\\Al$_{0.75}$GaAs(200nm)/GaAs(500nm)/GaAs(Sub).  Following arsenide growth, samples were cooled to below 350\textdegree C in an As$_2$ overpressure. Upon cooling, the surface reconstruction observed by reflection high energy electron diffraction (RHEED) changed from a (2x4)/c(2x8) observed during and immediately after growth to the expected As-rich c(4x4) surface reconstruction. Once the sample was cooled, the arsenic valve was closed, the sample was removed from the MBE chamber, and was further cooled to room temperature in the UHV prep chamber for a duration of at least 10 hours.

After cooling to room temperature, samples were loaded back in to the MBE growth chamber for epitaxial aluminum growth. Previous reports have shown that the orientation of Al thin films grown on GaAs(100) can be controlled using the GaAs surface termination, growth rate, and growth temperature\cite{Cho1978,Petroff1981,Ludeke1981,Pilkington1999}. When the substrate is kept at room temperature, Al grows predominantly in the (110) orientation on As-rich GaAs(100) and in the (100) orientation on the Ga-rich GaAs(100) surface. These results were repeated using the system employed for this study on separate calibration samples which were studied with X-ray diffraction (XRD) and atomic force microscopy (AFM). 

The wafer-bonding procedure used to complete the tri-layer process is very sensitive to surface roughness and therefore minimizing surface roughness of the epi-Al layer is required. The lowest surface roughness was observed for Al(100) grown on the Ga-rich GaAs(100) surface which was used for the remainder of the process for that reason. An As-rich GaAs(100) surface can be transformed to Ga-rich by either reducing the As-flux during growth or by annealing in the absence of, or in a reduced As overpressure. The surface reconstruction may be monitored by RHEED with the Ga-rich surface indicated by a c(8x2) or (4x6) surface reconstruction\cite{Cho1976,Ohtake2004}.

\begin{figure}
\includegraphics{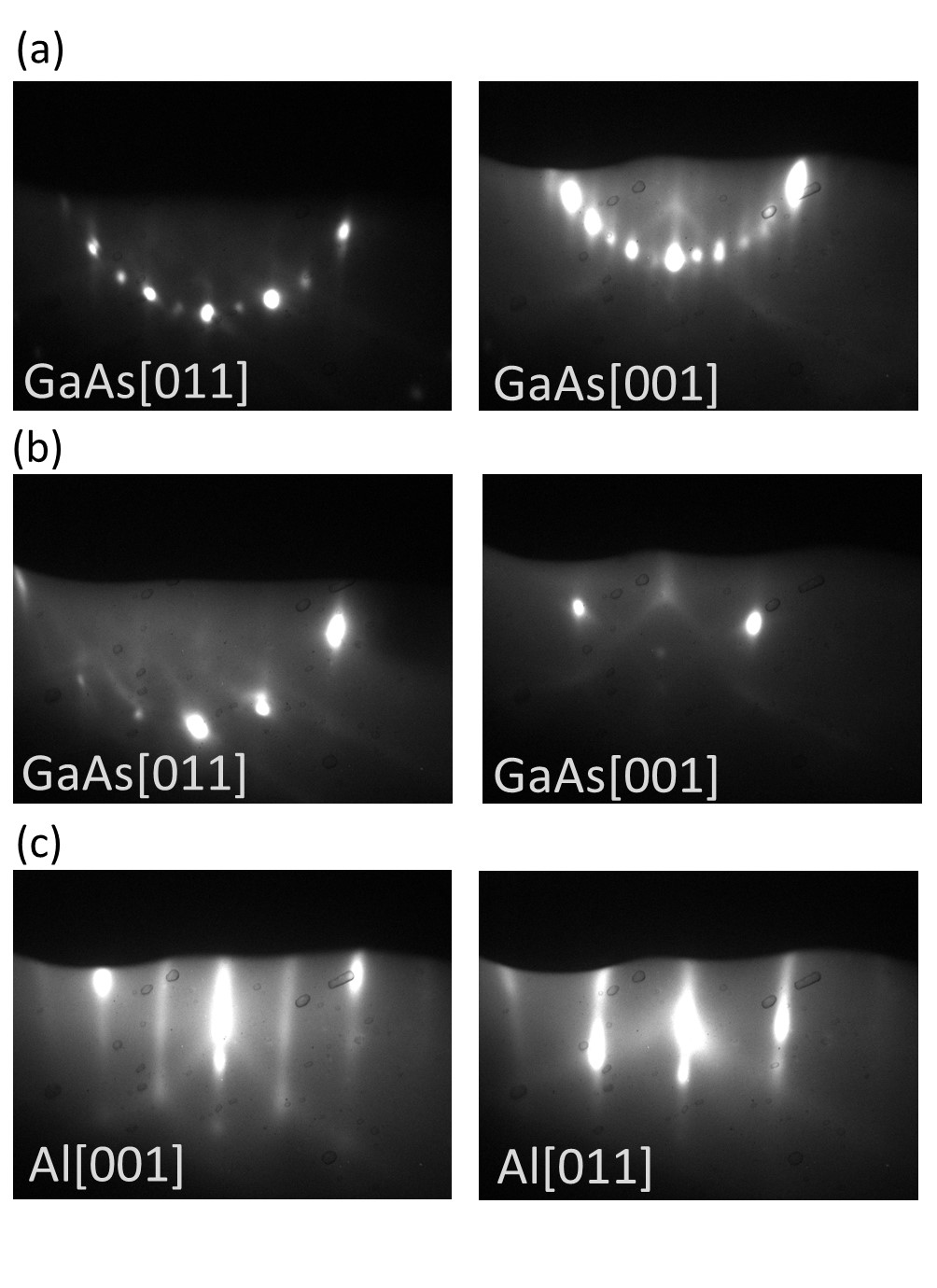}
\caption{\label{AlRHEED}Evolution of RHEED patterns observed during Al growth. (a) c(4x4) reconstructed GaAs at room temperature. (b) The same surface following deposition of 2ML Ga. (c) 40nm Al(100) deposited on the surface shown in (b) with the expected epitaxial relationship Al[001]||GaAs[011].}
\end{figure}

While obtaining a Ga-rich surface near growth temperature is straightforward, it was found that maintaining this Ga-rich surface while cooling the sample to below 350\textdegree C was difficult to reproduce due to residual arsenic partial pressure immediately following MBE growth. In order to achieve the required reproducibility, samples were cooled to room temperature with the As-rich c(4x4) reconstruction as was described previously. 2 ML of Ga was deposited on the c(4x4) surface at room temperature and the RHEED was observed to transform to a (1x1) or unreconstructed surface. Al was immediately grown on this surface at a rate of 0.8 \AA /sec. A 4-fold symmetric RHEED pattern consistent with Al(100) was observed which was confirmed by XRD. The evolution of the RHEED patterns during this procedure are included as Fig.~\ref{AlRHEED}.

\section{Wafer Bonding and Substrate Removal}
After Al growth, samples were removed from vacuum, coated with a layer of photoresist which is used to protect the wafer surface from particles, and cleaved into quarters of the original 2 inch diameter wafers. Photoresist was then stripped in acetone, rinsed with isopropanol (ACE/ISO), a thin 5nm Ti layer was deposited by e-beam evaporation to serve as an adhesion layer, and 20nm of aluminum oxide (AlO$_x$) was deposited by atomic layer deposition (ALD) using TMA and H$_2$O at a substrate temperature of 300\textdegree C. After AlO$_x$ deposition, samples were again cleaned in acetone along with a 3-inch diameter as-received Al$_2$O$_3$(0001) wafer for bonding. After rinsing in isopropanol, both the AlO$_x$/Al/GaAs sample and the sapphire bonding wafer were immediately loaded into an EVG 810 plasma activation tool. After pumping down, the wafers were exposed to an O$_2$ plasma for 30 seconds to activate the surface. 

Following the surface activation process, samples were exposed to atmosphere, the surfaces were immediately brought into contact, and pressed together using graphite clamping fixtures. Pressure was administered by applying a calibrated torque to nuts on the press fixture resulting in an approximate pressure of 200kPa. The press fixtures containing the samples were then loaded into an air oven held at 300\textdegree C for a period of about 12 hours to complete the wafer-bonding process. With the exception of the Ti and AlO$_x$ coating layers, this wafer bonding procedure is very similar to the low temperature plasma assisted bonding procedure used for InP/Si direct bonding which is discussed in more detail by Pasquariello and Hjart\cite{Pasquariello2002}.

After wafer bonding was completed, the GaAs(100) substrates were mechanically thinned prior to selective wet etching. The Sapphire bonding wafer was attached to a polishing fixture using Crystalbond wax followed by mechanical polishing using 600 grit sandpaper with isopropanol as a lubricant. The GaAs substrate was thinned using this technique from the initial wafer thickness of 350 micron to approximately 150 micron. After mechanical polishing, the sample was removed from the polishing fixture and cleaned with ultrasonic agitation in acetone in two-steps to remove the adhesive and particles created during the mechanical thinning. The GaAs substrate was then removed completely using H$_2$O$_2$:NH$_4$OH which can be a highly selective etchant of GaAs with respect to AlGaAs\cite{Collins1997}. 

A three step etching procedure was used with varying ratios of H$_2$O$_2$:NH$_4$OH starting with 12:1 for 15 minutes followed by 30:1 for approximately 20 minutes longer until the GaAs substrate was entirely removed revealing the first 200nm thick AlGaAs etch stop layer as indicated by a shiny surface. After the GaAs substrate was removed, the sample was transferred to a third H$_2$O$_2$:NH$_4$OH solution having concentration 40:1 for 5 minutes to remove residual GaAs substrate not visible by eye. A stir bar was used in all of the etchant solutions and the wafer placement in the beakers ensured that solution was constantly flowing over the entire wafer surface. The three-step etch was found to result in a more reproducible procedure.

Once the substrate was removed, the remaining MBE grown etch stop layers were selectively removed. Buffered HF (BHF) is a highly selective etch of Al$_x$GaAs with respect to GaAs for x>0.7\cite{Kim1998} and was used to remove the Al$_{0.75}$GaAs layers. Following removal of the first 200nm thick AlGaAs layer, the 250nm GaAs etch stop was removed using a 10:1 1M C$_6$H$_8$O$_7$:H$_2$O$_2$ solution which is another known selective etchant for GaAs with respect to AlGaAs\cite{Kim1998}. The citric acid solution was chosen over the H$_2$O$_2$:NH$_4$OH solution used to selectively remove the GaAs substrate as it was found to be more controllable for removal of thin GaAs films. The final AlGaAs etch stop layer was then removed using BHF, and the sample was immediately loaded in to the MBE growth system for surface cleaning and aluminum regrowth.

\section{Surface Cleaning and Aluminum Regrowth}

If both Al/GaAs interfaces in the tri-layer structure are to be pristine, the native oxide formed on the sample surface by exposure to atmosphere must be removed. The native oxide formed on the sample surfaces was removed at temperatures below 400\textdegree C using atomic hydrogen. This low temperature process was used as opposed to thermal desorption both to protect the wafer bond, and to maintain the pristine Al/GaAs interface. In order to determine appropriate conditions for native oxide removal on the GaAs surfaces resulting from the substrate removal and selective etch processes, three samples were prepared in the manner just described and loaded in to an ultrahigh vacuum (UHV) growth and characterization cluster tool equipped with atomic hydrogen cleaning, RHEED, and X-ray photoelectron spectroscopy (XPS) which was used to study the surface structure and chemistry. 

\begin{figure}
\includegraphics{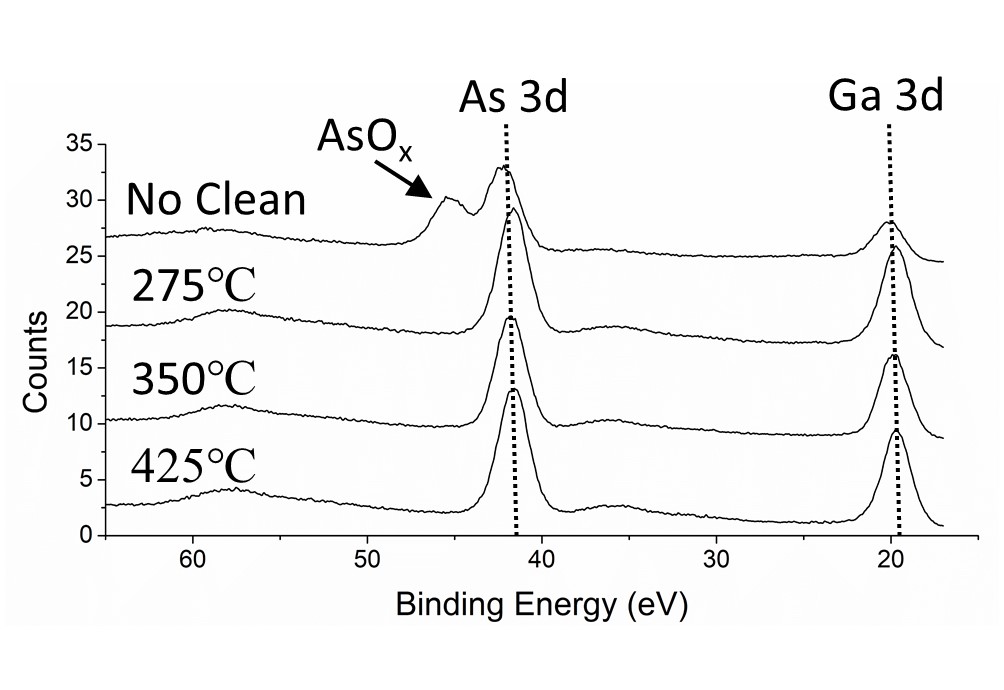}
\caption{\label{XPS}XPS measurements of prepared GaAs surfaces following hydrogen cleaning at variable temperatures. The vanishing of the high binding energy shoulder on the As 3d peak indicates reduction of the surface oxide.}
\end{figure}

RHEED and XPS measurements taken before hydrogen cleaning confirm the presence of an amorphous oxide on the surface indicated by a diffuse RHEED pattern along with the presence of a prominent oxygen 1s peak observed in XPS. After RHEED and XPS measurements, samples were sequentially exposed to an atomic hydrogen flux provided by a thermal cracker source operated at 1700\textdegree C. The hydrogen flux was kept the same for all samples which was controlled by adjusting a leak valve to obtain a set pressure of 1e-6 Torr in a chamber having a base pressure <1e-9 Torr. Samples were cleaned for 1 hour at variable temperatures of 275, 350, and 425\textdegree C. XPS measurements taken before and after hydrogen cleaning are included as Fig.~\ref{XPS}.

A clear reduction of the O 1s (not shown) as well as the high binding energy shoulder on the As 3d XPS peaks indicates removal of surface oxides by hydrogen cleaning. Similar XPS scans were observed for all temperatures tested, which demonstrates a large substrate temperature window for oxide removal using atomic hydrogen in this case. RHEED measurements transform from a diffuse background before oxide removal to a bright and clear diffraction pattern after oxide removal indicating a single crystal surface following hydrogen cleaning. 

\begin{figure}
\includegraphics{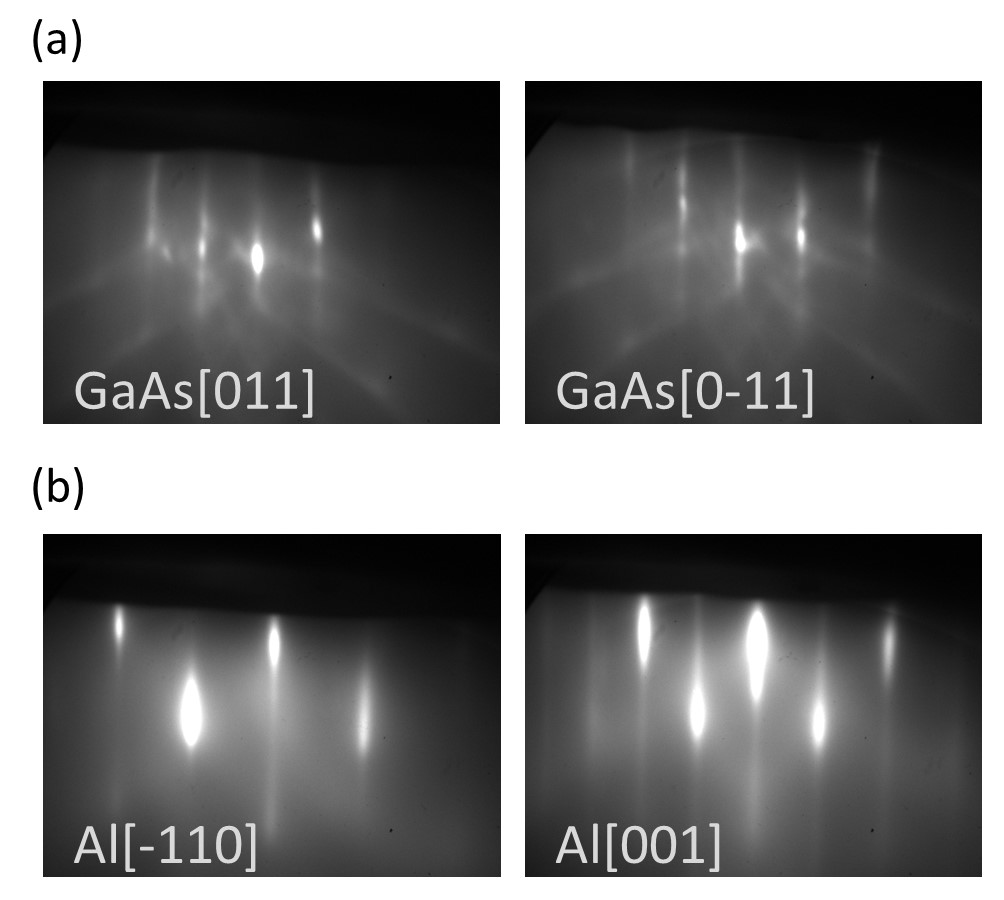}
\caption{\label{RegrowthRHEED}RHEED patterns of the wafer bonded GaAs(100) surface (a) following atomic hydrogen cleaning and (b) after growth of 50nm (110)Al showing the expected epitaxial relationship Al[-110]||GaAs[011].}
\end{figure}

In order to complete the tri-layer process, samples were hydrogen cleaned at a substrate temperature near 300\textdegree C using a hydrogen flux similar to that just described. Following hydrogen cleaning, samples were allowed to cool to room temperature for > 4 hours and then loaded in to an MBE chamber for Al regrowth. 50nm of Al was grown at room temperature at a rate of 0.8\AA /sec, completing the Al/GaAs/Al structure. RHEED was observed before and during Al growth. RHEED of the GaAs(100) following hydrogen cleaning resulted in a (1x1) RHEED pattern. The 2-fold symmetric RHEED pattern of the Al was found to be consistent with Al(110) having a single in-plane rotational domain. RHEED images before and after Al regrowth are shown in Fig.~\ref{RegrowthRHEED}.

\section{Structural Characterization of tri-layers}
\begin{figure}
\includegraphics{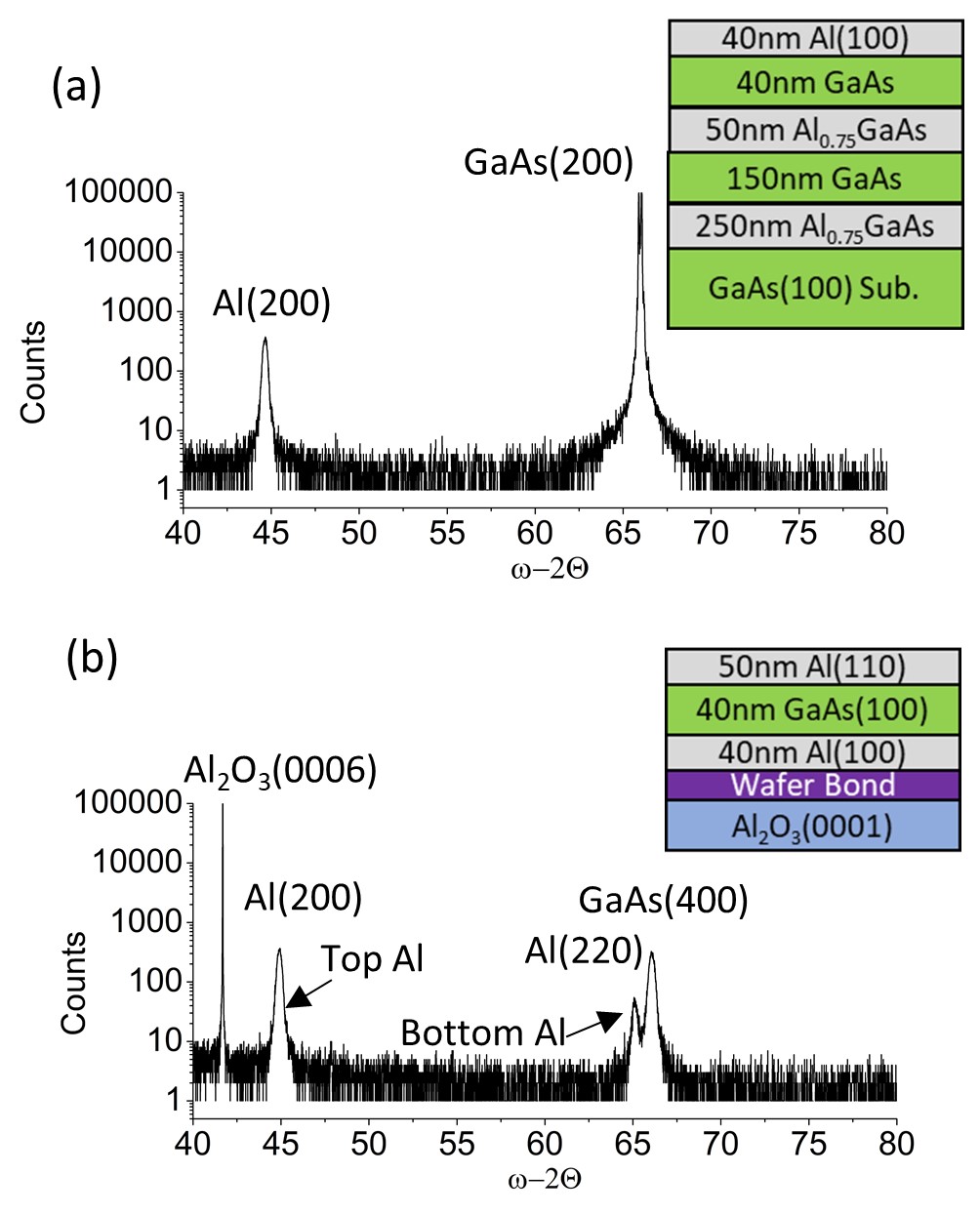}
\caption{\label{XRD} XRD measurement (a) before and (b) after the wafer bonding and regrowth process along with layer schematics of the structures.}
\end{figure}

X-ray diffraction (XRD) was performed before and after the wafer-bonding/Al regrowth process. Data taken from a sample having a 40nm thick GaAs layer is shown in Fig.~\ref{XRD}. 
The omega-2theta measurement taken before wafer bonding shown in Fig.~\ref{XRD}(a) shows prominent peaks from the GaAs substrate and the AlGaAs etch stop layers and confirms that the first Al-layer is (100) oriented indicated by the presence of an (200) Al diffraction peak and the absence of any other peaks. Figure~\ref{XRD}(b) shows the same measurement taken on the same sample after the wafer-bonding and Al-regrowth process. There are three prominent differences between the two measurements: the appearance of the Al$_2$O$_3$(0006) peak from the substrate that the tri-layer is bonded to, the reduction of the GaAs substrate peak to a smaller peak from the 40nm thick GaAs layer, and the appearance of an Al(220) peak from the regrown Al epi-layer.

\begin{figure*}
\includegraphics{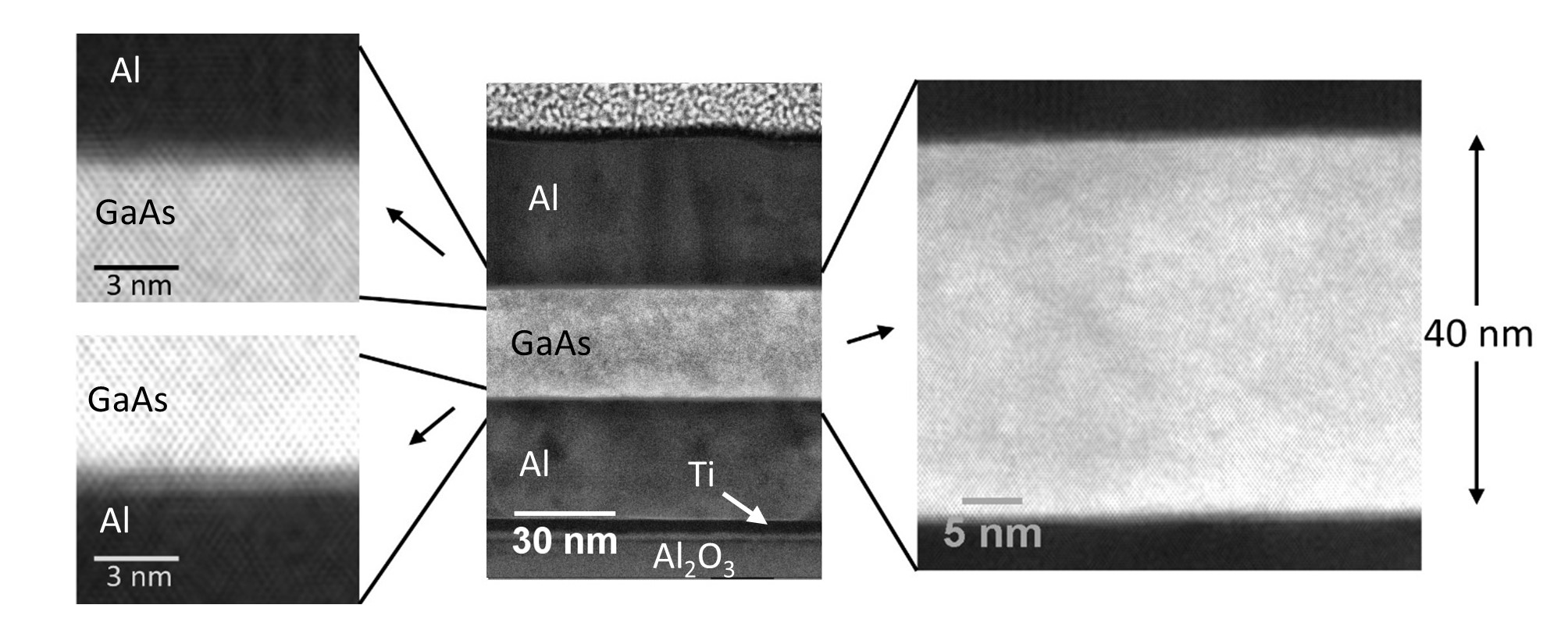}
\caption{\label{TEM}TEM images of the wafer bonded tri-layer structure and Al/GaAs interfaces. Both interfaces are abrupt and epitaxial and the GaAs is crystalline throughout its thickness.}
\end{figure*}

Cross-section high-angle annular dark field scanning transmission electron microscopy (HAADF-STEM) was performed on the same Al/GaAs/Al tri-layer structure having a 40nm thick GaAs layer and images are included as Fig.~\ref{TEM}. The images show that the GaAs layer is structurally uniform and crystalline throughout its thickness. Both Al/GaAs interfaces are observed to be abrupt and epitaxial.

\section{Discussion}
Among the potential applications for epitaxial super/semi/super tri-layers is their use in quantum information processing applications such as employing tri-layer capacitors and Josephson junctions in transmon circuits. Epitaxial tri-layers could replace large area planar capacitors and AlO$_x$ based JJs used in transmons, increasing scalability and coherence times, but the loss associated with the interfaces as well as the semiconductor itself need to be quite low in order to improve upon current transmon technology. This work has demonstrated that clean epitaxial interfaces may be formed which are expected to result in low loss. The loss inherent to the GaAs itself, however is not known and the obvious next step following this work is to measure the loss tangent and transport properties of the tri-layers and JJs.

The intrinsic loss tangent of III-V semiconductors including GaAs is expected to be higher than elemental semiconductors such as Si and Ge owing to their piezoelectricity as was discussed by Casparis et.al.\cite{Casparis2018} for InP(100) substrates and Scigliuzzo et al. \cite{Scigliuzzo2020} for GaAs substrates. Thus, super/semi/super tri-layers utilizing Si or Ge are expected to show superior performance for this application as compared to GaAs or other III-V materials. Though this work has presented a process for GaAs-based tri-layers, it is expected that this process may be extended to Si-based structures. Si epi-layers could be grown on III-V substrates and etch stop layers such as the lattice matched GaP/AlGaP system. Much of the remaining process would remain the same and similar selective wet etchants may be used.

Extending the wafer bonding process to semiconductor heterostructures is another potential application. The voltage-tunable gatemon qubits presented by Casparis et al\cite{Casparis2018} were fabricated using InAs quantum well heterostructures grown on InP(100) substrates. While the initial results are promising, the remaining qubit components such as resonators and readout cavities were deposited directly on the lossy InP substrate which limited performance. These limitations could be overcome by bonding structures to low loss substrates such as sapphire or silicon with epitaxial superconductor contacts potentially being regrown after wafer bonding.

\section{Summary}
Epitaxial Al/GaAs/Al tri-layer structures have been made using a novel wafer-bonding and regrowth technique. Because the GaAs layer is initially grown on AlGaAs using standard MBE growth procedures, it is high quality semiconductor that can be made with arbitrary thickness. The crystal orientation of both aluminum layers may be controlled by GaAs(100) surface termination and growth temperature where (100), (110), and (111)Al all may be grown on GaAs(100). The presented process may be extended to other semiconductors and semiconductor heterostructures which could be used to improve transmon and gatemon technology beyond state-of-the-art.

\begin{acknowledgments}
This research was funded through the new and emerging qubit science and technology (NEQST) program initiated by the US Army Research Office (ARO) in collaboration with the Laboratory of Physical Sciences (LPS) under grant W911NF1810114. A portion of this work was performed in the UCSB Nanofabrication Facility, an open access laboratory. XRD measurements were conducted in the UCSB materials research lab (MRL). The MRL Shared Experimental Facilities are supported by the MRSEC Program of the NSF under Award No. DMR 1720256; a member of the NSF-funded Materials Research Facilities Network.
\end{acknowledgments}

\nocite{*}
\bibliography{aipsamp}

\providecommand{\noopsort}[1]{}\providecommand{\singleletter}[1]{#1}%
\begin{thebibliography}{39}%
\makeatletter
\providecommand \@ifxundefined [1]{%
 \@ifx{#1\undefined}
}%
\providecommand \@ifnum [1]{%
 \ifnum #1\expandafter \@firstoftwo
 \else \expandafter \@secondoftwo
 \fi
}%
\providecommand \@ifx [1]{%
 \ifx #1\expandafter \@firstoftwo
 \else \expandafter \@secondoftwo
 \fi
}%
\providecommand \natexlab [1]{#1}%
\providecommand \enquote  [1]{``#1''}%
\providecommand \bibnamefont  [1]{#1}%
\providecommand \bibfnamefont [1]{#1}%
\providecommand \citenamefont [1]{#1}%
\providecommand \href@noop [0]{\@secondoftwo}%
\providecommand \href [0]{\begingroup \@sanitize@url \@href}%
\providecommand \@href[1]{\@@startlink{#1}\@@href}%
\providecommand \@@href[1]{\endgroup#1\@@endlink}%
\providecommand \@sanitize@url [0]{\catcode `\\12\catcode `\$12\catcode
  `\&12\catcode `\#12\catcode `\^12\catcode `\_12\catcode `\%12\relax}%
\providecommand \@@startlink[1]{}%
\providecommand \@@endlink[0]{}%
\providecommand \url  [0]{\begingroup\@sanitize@url \@url }%
\providecommand \@url [1]{\endgroup\@href {#1}{\urlprefix }}%
\providecommand \urlprefix  [0]{URL }%
\providecommand \Eprint [0]{\href }%
\providecommand \doibase [0]{http://dx.doi.org/}%
\providecommand \selectlanguage [0]{\@gobble}%
\providecommand \bibinfo  [0]{\@secondoftwo}%
\providecommand \bibfield  [0]{\@secondoftwo}%
\providecommand \translation [1]{[#1]}%
\providecommand \BibitemOpen [0]{}%
\providecommand \bibitemStop [0]{}%
\providecommand \bibitemNoStop [0]{.\EOS\space}%
\providecommand \EOS [0]{\spacefactor3000\relax}%
\providecommand \BibitemShut  [1]{\csname bibitem#1\endcsname}%
\let\auto@bib@innerbib\@empty
\bibitem [{\citenamefont {Krantz}\ \emph {et~al.}(2019)\citenamefont {Krantz},
  \citenamefont {Kjaergaard}, \citenamefont {Yan}, \citenamefont {Orlando},
  \citenamefont {Gustavsson},\ and\ \citenamefont {Oliver}}]{Krantz2019}%
  \BibitemOpen
  \bibfield  {author} {\bibinfo {author} {\bibfnamefont {P.}~\bibnamefont
  {Krantz}}, \bibinfo {author} {\bibfnamefont {M.}~\bibnamefont {Kjaergaard}},
  \bibinfo {author} {\bibfnamefont {F.}~\bibnamefont {Yan}}, \bibinfo {author}
  {\bibfnamefont {T.}~\bibnamefont {Orlando}}, \bibinfo {author} {\bibfnamefont
  {S.}~\bibnamefont {Gustavsson}}, \ and\ \bibinfo {author} {\bibfnamefont
  {W.}~\bibnamefont {Oliver}},\ }\href@noop {} {\bibfield  {journal} {\bibinfo
  {journal} {Appl. Phys. Rev.}\ }\textbf {\bibinfo {volume} {6}},\ \bibinfo
  {pages} {021318} (\bibinfo {year} {2019})}\BibitemShut {NoStop}%
\bibitem [{\citenamefont {Wendin}(2017)}]{Wendin2017}%
  \BibitemOpen
  \bibfield  {author} {\bibinfo {author} {\bibfnamefont {G.}~\bibnamefont
  {Wendin}},\ }\href@noop {} {\bibfield  {journal} {\bibinfo  {journal} {Rep.
  Prog. Phys.}\ }\textbf {\bibinfo {volume} {80}},\ \bibinfo {pages} {106001}
  (\bibinfo {year} {2017})}\BibitemShut {NoStop}%
\bibitem [{\citenamefont {Larsen}\ \emph {et~al.}(2018)\citenamefont {Larsen},
  \citenamefont {Peterson}, \citenamefont {Kuemmeth}, \citenamefont
  {Jesperson}, \citenamefont {Krogstrup}, \citenamefont {Nygard},\ and\
  \citenamefont {Marcus}}]{Larsen2015}%
  \BibitemOpen
  \bibfield  {author} {\bibinfo {author} {\bibfnamefont {T.}~\bibnamefont
  {Larsen}}, \bibinfo {author} {\bibfnamefont {K.}~\bibnamefont {Peterson}},
  \bibinfo {author} {\bibfnamefont {F.}~\bibnamefont {Kuemmeth}}, \bibinfo
  {author} {\bibfnamefont {T.}~\bibnamefont {Jesperson}}, \bibinfo {author}
  {\bibfnamefont {P.}~\bibnamefont {Krogstrup}}, \bibinfo {author}
  {\bibfnamefont {J.}~\bibnamefont {Nygard}}, \ and\ \bibinfo {author}
  {\bibfnamefont {C.}~\bibnamefont {Marcus}},\ }\href@noop {} {\bibfield
  {journal} {\bibinfo  {journal} {Nature Nanotech.}\ }\textbf {\bibinfo
  {volume} {13}},\ \bibinfo {pages} {915--919} (\bibinfo {year}
  {2018})}\BibitemShut {NoStop}%
\bibitem [{\citenamefont {Casparis}\ \emph {et~al.}(2018)\citenamefont
  {Casparis}, \citenamefont {Connolly}, \citenamefont {Kjaergaard},
  \citenamefont {Pearson}, \citenamefont {Kringhoj}, \citenamefont {Larsen},
  \citenamefont {Kuemmeth}, \citenamefont {Wang}, \citenamefont {Thomas},
  \citenamefont {Gronin}, \citenamefont {Gardner}, \citenamefont {Manfra},
  \citenamefont {Marcus},\ and\ \citenamefont {Peterson}}]{Casparis2018}%
  \BibitemOpen
  \bibfield  {author} {\bibinfo {author} {\bibfnamefont {L.}~\bibnamefont
  {Casparis}}, \bibinfo {author} {\bibfnamefont {M.}~\bibnamefont {Connolly}},
  \bibinfo {author} {\bibfnamefont {M.}~\bibnamefont {Kjaergaard}}, \bibinfo
  {author} {\bibfnamefont {N.}~\bibnamefont {Pearson}}, \bibinfo {author}
  {\bibfnamefont {A.}~\bibnamefont {Kringhoj}}, \bibinfo {author}
  {\bibfnamefont {T.}~\bibnamefont {Larsen}}, \bibinfo {author} {\bibfnamefont
  {F.}~\bibnamefont {Kuemmeth}}, \bibinfo {author} {\bibfnamefont
  {T.}~\bibnamefont {Wang}}, \bibinfo {author} {\bibfnamefont {C.}~\bibnamefont
  {Thomas}}, \bibinfo {author} {\bibfnamefont {S.}~\bibnamefont {Gronin}},
  \bibinfo {author} {\bibfnamefont {G.}~\bibnamefont {Gardner}}, \bibinfo
  {author} {\bibfnamefont {M.}~\bibnamefont {Manfra}}, \bibinfo {author}
  {\bibfnamefont {C.}~\bibnamefont {Marcus}}, \ and\ \bibinfo {author}
  {\bibfnamefont {K.}~\bibnamefont {Peterson}},\ }\href@noop {} {\bibfield
  {journal} {\bibinfo  {journal} {Nature Nanotech.}\ }\textbf {\bibinfo
  {volume} {13}},\ \bibinfo {pages} {915--919} (\bibinfo {year}
  {2018})}\BibitemShut {NoStop}%
\bibitem [{\citenamefont {Mourik}\ \emph {et~al.}(2012)\citenamefont {Mourik},
  \citenamefont {Zuo}, \citenamefont {Frolov}, \citenamefont {Plissard},
  \citenamefont {Bakkers},\ and\ \citenamefont {Kouwenhoven}}]{Mourik2012}%
  \BibitemOpen
  \bibfield  {author} {\bibinfo {author} {\bibfnamefont {V.}~\bibnamefont
  {Mourik}}, \bibinfo {author} {\bibfnamefont {K.}~\bibnamefont {Zuo}},
  \bibinfo {author} {\bibfnamefont {S.}~\bibnamefont {Frolov}}, \bibinfo
  {author} {\bibfnamefont {S.}~\bibnamefont {Plissard}}, \bibinfo {author}
  {\bibfnamefont {E.}~\bibnamefont {Bakkers}}, \ and\ \bibinfo {author}
  {\bibfnamefont {L.}~\bibnamefont {Kouwenhoven}},\ }\href@noop {} {\bibfield
  {journal} {\bibinfo  {journal} {Science}\ }\textbf {\bibinfo {volume}
  {336}},\ \bibinfo {pages} {1003--1007} (\bibinfo {year} {2012})}\BibitemShut
  {NoStop}%
\bibitem [{\citenamefont {Shabani}\ \emph {et~al.}(2016)\citenamefont
  {Shabani}, \citenamefont {Kjaergaard}, \citenamefont {Suominen},
  \citenamefont {Kim}, \citenamefont {Nichele}, \citenamefont {Paksouski},
  \citenamefont {Stankevic}, \citenamefont {Lutchyn}, \citenamefont
  {Krogstrup}, \citenamefont {Feidenhans'l}, \citenamefont {Kraemer},
  \citenamefont {Nayak}, \citenamefont {Troyer}, \citenamefont {Marcus},\ and\
  \citenamefont {Palmstrom}}]{Shabani2016}%
  \BibitemOpen
  \bibfield  {author} {\bibinfo {author} {\bibfnamefont {J.}~\bibnamefont
  {Shabani}}, \bibinfo {author} {\bibfnamefont {M.}~\bibnamefont {Kjaergaard}},
  \bibinfo {author} {\bibfnamefont {H.}~\bibnamefont {Suominen}}, \bibinfo
  {author} {\bibfnamefont {Y.}~\bibnamefont {Kim}}, \bibinfo {author}
  {\bibfnamefont {F.}~\bibnamefont {Nichele}}, \bibinfo {author} {\bibfnamefont
  {K.}~\bibnamefont {Paksouski}}, \bibinfo {author} {\bibfnamefont
  {T.}~\bibnamefont {Stankevic}}, \bibinfo {author} {\bibfnamefont
  {R.}~\bibnamefont {Lutchyn}}, \bibinfo {author} {\bibfnamefont
  {P.}~\bibnamefont {Krogstrup}}, \bibinfo {author} {\bibfnamefont
  {R.}~\bibnamefont {Feidenhans'l}}, \bibinfo {author} {\bibfnamefont
  {S.}~\bibnamefont {Kraemer}}, \bibinfo {author} {\bibfnamefont
  {C.}~\bibnamefont {Nayak}}, \bibinfo {author} {\bibfnamefont
  {M.}~\bibnamefont {Troyer}}, \bibinfo {author} {\bibfnamefont
  {C.}~\bibnamefont {Marcus}}, \ and\ \bibinfo {author} {\bibfnamefont
  {C.}~\bibnamefont {Palmstrom}},\ }\href@noop {} {\bibfield  {journal}
  {\bibinfo  {journal} {Phys. Rev. B}\ }\textbf {\bibinfo {volume} {93}},\
  \bibinfo {pages} {155402} (\bibinfo {year} {2016})}\BibitemShut {NoStop}%
\bibitem [{\citenamefont {Karzig}\ \emph {et~al.}(2017)\citenamefont {Karzig},
  \citenamefont {Knapp}, \citenamefont {Lutchyn}, \citenamefont {Bonderson},
  \citenamefont {Hastings}, \citenamefont {Nayak}, \citenamefont {Alicea},
  \citenamefont {Flensberg}, \citenamefont {Plugge}, \citenamefont {Oreg},
  \citenamefont {Marcus},\ and\ \citenamefont {Freedman}}]{Karzig2017}%
  \BibitemOpen
  \bibfield  {author} {\bibinfo {author} {\bibfnamefont {T.}~\bibnamefont
  {Karzig}}, \bibinfo {author} {\bibfnamefont {C.}~\bibnamefont {Knapp}},
  \bibinfo {author} {\bibfnamefont {R.}~\bibnamefont {Lutchyn}}, \bibinfo
  {author} {\bibfnamefont {P.}~\bibnamefont {Bonderson}}, \bibinfo {author}
  {\bibfnamefont {M.}~\bibnamefont {Hastings}}, \bibinfo {author}
  {\bibfnamefont {C.}~\bibnamefont {Nayak}}, \bibinfo {author} {\bibfnamefont
  {J.}~\bibnamefont {Alicea}}, \bibinfo {author} {\bibfnamefont
  {K.}~\bibnamefont {Flensberg}}, \bibinfo {author} {\bibfnamefont
  {S.}~\bibnamefont {Plugge}}, \bibinfo {author} {\bibfnamefont
  {Y.}~\bibnamefont {Oreg}}, \bibinfo {author} {\bibfnamefont {C.}~\bibnamefont
  {Marcus}}, \ and\ \bibinfo {author} {\bibfnamefont {M.}~\bibnamefont
  {Freedman}},\ }\href@noop {} {\bibfield  {journal} {\bibinfo  {journal}
  {Phys. Rev. B}\ }\textbf {\bibinfo {volume} {95}},\ \bibinfo {pages} {235305}
  (\bibinfo {year} {2017})}\BibitemShut {NoStop}%
\bibitem [{\citenamefont {Wisbey}\ \emph {et~al.}(2010)\citenamefont {Wisbey},
  \citenamefont {Gao}, \citenamefont {Vissers}, \citenamefont {da~Silva},
  \citenamefont {Kline}, \citenamefont {Vale},\ and\ \citenamefont
  {Pappas}}]{Wisbey2010}%
  \BibitemOpen
  \bibfield  {author} {\bibinfo {author} {\bibfnamefont {D.}~\bibnamefont
  {Wisbey}}, \bibinfo {author} {\bibfnamefont {J.}~\bibnamefont {Gao}},
  \bibinfo {author} {\bibfnamefont {M.}~\bibnamefont {Vissers}}, \bibinfo
  {author} {\bibfnamefont {F.}~\bibnamefont {da~Silva}}, \bibinfo {author}
  {\bibfnamefont {J.}~\bibnamefont {Kline}}, \bibinfo {author} {\bibfnamefont
  {L.}~\bibnamefont {Vale}}, \ and\ \bibinfo {author} {\bibfnamefont
  {D.}~\bibnamefont {Pappas}},\ }\href@noop {} {\bibfield  {journal} {\bibinfo
  {journal} {J. Appl. Phys.}\ }\textbf {\bibinfo {volume} {108}},\ \bibinfo
  {pages} {093918} (\bibinfo {year} {2010})}\BibitemShut {NoStop}%
\bibitem [{\citenamefont {Vissers}\ \emph {et~al.}(2012)\citenamefont
  {Vissers}, \citenamefont {Weides}, \citenamefont {Kline}, \citenamefont
  {Sandberg},\ and\ \citenamefont {Pappas}}]{Vissers2012}%
  \BibitemOpen
  \bibfield  {author} {\bibinfo {author} {\bibfnamefont {M.}~\bibnamefont
  {Vissers}}, \bibinfo {author} {\bibfnamefont {M.}~\bibnamefont {Weides}},
  \bibinfo {author} {\bibfnamefont {J.}~\bibnamefont {Kline}}, \bibinfo
  {author} {\bibfnamefont {M.}~\bibnamefont {Sandberg}}, \ and\ \bibinfo
  {author} {\bibfnamefont {D.}~\bibnamefont {Pappas}},\ }\href@noop {}
  {\bibfield  {journal} {\bibinfo  {journal} {Appl. Phys. Lett.}\ }\textbf
  {\bibinfo {volume} {101}},\ \bibinfo {pages} {022601} (\bibinfo {year}
  {2012})}\BibitemShut {NoStop}%
\bibitem [{\citenamefont {Wang}\ \emph {et~al.}(2015)\citenamefont {Wang},
  \citenamefont {Axline}, \citenamefont {Gao}, \citenamefont {Brecht},
  \citenamefont {Chu}, \citenamefont {Frunzio}, \citenamefont {Devoret},\ and\
  \citenamefont {Rchoelkopf}}]{Wang2015}%
  \BibitemOpen
  \bibfield  {author} {\bibinfo {author} {\bibfnamefont {C.}~\bibnamefont
  {Wang}}, \bibinfo {author} {\bibfnamefont {C.}~\bibnamefont {Axline}},
  \bibinfo {author} {\bibfnamefont {Y.}~\bibnamefont {Gao}}, \bibinfo {author}
  {\bibfnamefont {T.}~\bibnamefont {Brecht}}, \bibinfo {author} {\bibfnamefont
  {Y.}~\bibnamefont {Chu}}, \bibinfo {author} {\bibfnamefont {L.}~\bibnamefont
  {Frunzio}}, \bibinfo {author} {\bibfnamefont {M.}~\bibnamefont {Devoret}}, \
  and\ \bibinfo {author} {\bibfnamefont {R.}~\bibnamefont {Rchoelkopf}},\
  }\href@noop {} {\bibfield  {journal} {\bibinfo  {journal} {Appl. Phys.
  Lett.}\ }\textbf {\bibinfo {volume} {107}},\ \bibinfo {pages} {162601}
  (\bibinfo {year} {2015})}\BibitemShut {NoStop}%
\bibitem [{\citenamefont {Woods}\ \emph {et~al.}(2019)\citenamefont {Woods},
  \citenamefont {Calusine}, \citenamefont {Melville}, \citenamefont {Sevi},
  \citenamefont {Golden}, \citenamefont {Kim}, \citenamefont {Rosenberg},
  \citenamefont {Yoder},\ and\ \citenamefont {Oliver}}]{Woods2019}%
  \BibitemOpen
  \bibfield  {author} {\bibinfo {author} {\bibfnamefont {W.}~\bibnamefont
  {Woods}}, \bibinfo {author} {\bibfnamefont {G.}~\bibnamefont {Calusine}},
  \bibinfo {author} {\bibfnamefont {A.}~\bibnamefont {Melville}}, \bibinfo
  {author} {\bibfnamefont {A.}~\bibnamefont {Sevi}}, \bibinfo {author}
  {\bibfnamefont {E.}~\bibnamefont {Golden}}, \bibinfo {author} {\bibfnamefont
  {D.}~\bibnamefont {Kim}}, \bibinfo {author} {\bibfnamefont {D.}~\bibnamefont
  {Rosenberg}}, \bibinfo {author} {\bibfnamefont {J.}~\bibnamefont {Yoder}}, \
  and\ \bibinfo {author} {\bibfnamefont {W.}~\bibnamefont {Oliver}},\
  }\href@noop {} {\bibfield  {journal} {\bibinfo  {journal} {Phys. Rev.
  Applied}\ }\textbf {\bibinfo {volume} {12}},\ \bibinfo {pages} {014012}
  (\bibinfo {year} {2019})}\BibitemShut {NoStop}%
\bibitem [{\citenamefont {Gambetta}\ \emph {et~al.}(2017)\citenamefont
  {Gambetta}, \citenamefont {Murray}, \citenamefont {Fung}, \citenamefont
  {McClure}, \citenamefont {Dial}, \citenamefont {Shanks}, \citenamefont
  {Sleight},\ and\ \citenamefont {Steffen}}]{Gambetta2017}%
  \BibitemOpen
  \bibfield  {author} {\bibinfo {author} {\bibfnamefont {J.}~\bibnamefont
  {Gambetta}}, \bibinfo {author} {\bibfnamefont {C.}~\bibnamefont {Murray}},
  \bibinfo {author} {\bibfnamefont {Y.}~\bibnamefont {Fung}}, \bibinfo {author}
  {\bibfnamefont {D.}~\bibnamefont {McClure}}, \bibinfo {author} {\bibfnamefont
  {O.}~\bibnamefont {Dial}}, \bibinfo {author} {\bibfnamefont {W.}~\bibnamefont
  {Shanks}}, \bibinfo {author} {\bibfnamefont {J.}~\bibnamefont {Sleight}}, \
  and\ \bibinfo {author} {\bibfnamefont {M.}~\bibnamefont {Steffen}},\
  }\href@noop {} {\bibfield  {journal} {\bibinfo  {journal} {IEEE Trans. Appl.
  Supercond.}\ }\textbf {\bibinfo {volume} {27}},\ \bibinfo {pages} {1}
  (\bibinfo {year} {2017})}\BibitemShut {NoStop}%
\bibitem [{\citenamefont {Cheng}, \citenamefont {Lutchyn},\ and\ \citenamefont
  {Sarma}(2012)}]{Cheng2012}%
  \BibitemOpen
  \bibfield  {author} {\bibinfo {author} {\bibfnamefont {M.}~\bibnamefont
  {Cheng}}, \bibinfo {author} {\bibfnamefont {R.}~\bibnamefont {Lutchyn}}, \
  and\ \bibinfo {author} {\bibfnamefont {S.~D.}\ \bibnamefont {Sarma}},\
  }\href@noop {} {\bibfield  {journal} {\bibinfo  {journal} {Phys. Rev. B}\
  }\textbf {\bibinfo {volume} {85}},\ \bibinfo {pages} {165124} (\bibinfo
  {year} {2012})}\BibitemShut {NoStop}%
\bibitem [{\citenamefont {Rainis}\ and\ \citenamefont
  {Loss}(2012)}]{Rainis2012}%
  \BibitemOpen
  \bibfield  {author} {\bibinfo {author} {\bibfnamefont {D.}~\bibnamefont
  {Rainis}}\ and\ \bibinfo {author} {\bibfnamefont {D.}~\bibnamefont {Loss}},\
  }\href@noop {} {\bibfield  {journal} {\bibinfo  {journal} {Phys. Rev. B}\
  }\textbf {\bibinfo {volume} {85}},\ \bibinfo {pages} {174533} (\bibinfo
  {year} {2012})}\BibitemShut {NoStop}%
\bibitem [{\citenamefont {Chang}\ \emph {et~al.}(2015)\citenamefont {Chang},
  \citenamefont {Albrecht}, \citenamefont {Jesperson}, \citenamefont
  {Kuemmeth}, \citenamefont {Krogstrup}, \citenamefont {Nygard},\ and\
  \citenamefont {Marcus}}]{Chang2015}%
  \BibitemOpen
  \bibfield  {author} {\bibinfo {author} {\bibfnamefont {W.}~\bibnamefont
  {Chang}}, \bibinfo {author} {\bibfnamefont {S.}~\bibnamefont {Albrecht}},
  \bibinfo {author} {\bibfnamefont {T.}~\bibnamefont {Jesperson}}, \bibinfo
  {author} {\bibfnamefont {F.}~\bibnamefont {Kuemmeth}}, \bibinfo {author}
  {\bibfnamefont {P.}~\bibnamefont {Krogstrup}}, \bibinfo {author}
  {\bibfnamefont {J.}~\bibnamefont {Nygard}}, \ and\ \bibinfo {author}
  {\bibfnamefont {C.}~\bibnamefont {Marcus}},\ }\href@noop {} {\bibfield
  {journal} {\bibinfo  {journal} {Nature Nanotech.}\ }\textbf {\bibinfo
  {volume} {10}},\ \bibinfo {pages} {232--236} (\bibinfo {year}
  {2015})}\BibitemShut {NoStop}%
\bibitem [{\citenamefont {Kjaergaard}\ \emph {et~al.}(2016)\citenamefont
  {Kjaergaard}, \citenamefont {Nichele}, \citenamefont {Suominen},
  \citenamefont {Nowak}, \citenamefont {Wimmer}, \citenamefont {Akhermov},
  \citenamefont {Folk}, \citenamefont {Flensberg}, \citenamefont {Shabani},
  \citenamefont {Palmstrom},\ and\ \citenamefont {Marcus}}]{Kjaergaard2016}%
  \BibitemOpen
  \bibfield  {author} {\bibinfo {author} {\bibfnamefont {M.}~\bibnamefont
  {Kjaergaard}}, \bibinfo {author} {\bibfnamefont {F.}~\bibnamefont {Nichele}},
  \bibinfo {author} {\bibfnamefont {H.}~\bibnamefont {Suominen}}, \bibinfo
  {author} {\bibfnamefont {M.}~\bibnamefont {Nowak}}, \bibinfo {author}
  {\bibfnamefont {M.}~\bibnamefont {Wimmer}}, \bibinfo {author} {\bibfnamefont
  {A.}~\bibnamefont {Akhermov}}, \bibinfo {author} {\bibfnamefont
  {J.}~\bibnamefont {Folk}}, \bibinfo {author} {\bibfnamefont {F.}~\bibnamefont
  {Flensberg}}, \bibinfo {author} {\bibfnamefont {J.}~\bibnamefont {Shabani}},
  \bibinfo {author} {\bibfnamefont {C.}~\bibnamefont {Palmstrom}}, \ and\
  \bibinfo {author} {\bibfnamefont {C.}~\bibnamefont {Marcus}},\ }\href@noop {}
  {\bibfield  {journal} {\bibinfo  {journal} {Nature Comm.}\ }\textbf {\bibinfo
  {volume} {7}},\ \bibinfo {pages} {12841} (\bibinfo {year}
  {2016})}\BibitemShut {NoStop}%
\bibitem [{\citenamefont {Fritz}\ \emph {et~al.}(2019)\citenamefont {Fritz},
  \citenamefont {Radtke}, \citenamefont {Schneider}, \citenamefont {Weides},\
  and\ \citenamefont {Gerthsen}}]{Fritz2019}%
  \BibitemOpen
  \bibfield  {author} {\bibinfo {author} {\bibfnamefont {S.}~\bibnamefont
  {Fritz}}, \bibinfo {author} {\bibfnamefont {L.}~\bibnamefont {Radtke}},
  \bibinfo {author} {\bibfnamefont {R.}~\bibnamefont {Schneider}}, \bibinfo
  {author} {\bibfnamefont {M.}~\bibnamefont {Weides}}, \ and\ \bibinfo {author}
  {\bibfnamefont {G.}~\bibnamefont {Gerthsen}},\ }\href@noop {} {\bibfield
  {journal} {\bibinfo  {journal} {J. Appl. Phys.}\ }\textbf {\bibinfo {volume}
  {125}},\ \bibinfo {pages} {165301} (\bibinfo {year} {2019})}\BibitemShut
  {NoStop}%
\bibitem [{\citenamefont {Zeng}\ \emph {et~al.}(2015)\citenamefont {Zeng},
  \citenamefont {Nik}, \citenamefont {Greibe}, \citenamefont {Krantz},
  \citenamefont {Wilson}, \citenamefont {Delsing},\ and\ \citenamefont
  {Olsson}}]{Zeng2015}%
  \BibitemOpen
  \bibfield  {author} {\bibinfo {author} {\bibfnamefont {L.}~\bibnamefont
  {Zeng}}, \bibinfo {author} {\bibfnamefont {S.}~\bibnamefont {Nik}}, \bibinfo
  {author} {\bibfnamefont {T.}~\bibnamefont {Greibe}}, \bibinfo {author}
  {\bibfnamefont {P.}~\bibnamefont {Krantz}}, \bibinfo {author} {\bibfnamefont
  {C.}~\bibnamefont {Wilson}}, \bibinfo {author} {\bibfnamefont
  {P.}~\bibnamefont {Delsing}}, \ and\ \bibinfo {author} {\bibfnamefont
  {E.}~\bibnamefont {Olsson}},\ }\href@noop {} {\bibfield  {journal} {\bibinfo
  {journal} {J. Phys. D: Appl. Phys.}\ }\textbf {\bibinfo {volume} {48}},\
  \bibinfo {pages} {395308} (\bibinfo {year} {2015})}\BibitemShut {NoStop}%
\bibitem [{\citenamefont {Tahan}(2019)}]{Tahan2019}%
  \BibitemOpen
  \bibfield  {author} {\bibinfo {author} {\bibfnamefont {C.}~\bibnamefont
  {Tahan}},\ }\href@noop {} {\bibfield  {journal} {\bibinfo  {journal} {Nature
  Nanotech.}\ }\textbf {\bibinfo {volume} {14}},\ \bibinfo {pages} {102--103}
  (\bibinfo {year} {2019})}\BibitemShut {NoStop}%
\bibitem [{\citenamefont {Cho}\ and\ \citenamefont {Dernier}(1978)}]{Cho1978}%
  \BibitemOpen
  \bibfield  {author} {\bibinfo {author} {\bibfnamefont {A.}~\bibnamefont
  {Cho}}\ and\ \bibinfo {author} {\bibfnamefont {P.}~\bibnamefont {Dernier}},\
  }\href@noop {} {\bibfield  {journal} {\bibinfo  {journal} {J. Appl. Phys.}\
  }\textbf {\bibinfo {volume} {49}},\ \bibinfo {pages} {6} (\bibinfo {year}
  {1978})}\BibitemShut {NoStop}%
\bibitem [{\citenamefont {Petroff}\ \emph {et~al.}(1981)\citenamefont
  {Petroff}, \citenamefont {Feldman}, \citenamefont {Cho},\ and\ \citenamefont
  {Williams}}]{Petroff1981}%
  \BibitemOpen
  \bibfield  {author} {\bibinfo {author} {\bibfnamefont {P.}~\bibnamefont
  {Petroff}}, \bibinfo {author} {\bibfnamefont {L.}~\bibnamefont {Feldman}},
  \bibinfo {author} {\bibfnamefont {A.}~\bibnamefont {Cho}}, \ and\ \bibinfo
  {author} {\bibfnamefont {R.}~\bibnamefont {Williams}},\ }\href@noop {}
  {\bibfield  {journal} {\bibinfo  {journal} {J. Appl. Phys.}\ }\textbf
  {\bibinfo {volume} {52}},\ \bibinfo {pages} {7317} (\bibinfo {year}
  {1981})}\BibitemShut {NoStop}%
\bibitem [{\citenamefont {R.Ludeke}\ and\ \citenamefont
  {Landgren}(1981)}]{Ludeke1981}%
  \BibitemOpen
  \bibfield  {author} {\bibinfo {author} {\bibnamefont {R.Ludeke}}\ and\
  \bibinfo {author} {\bibfnamefont {G.}~\bibnamefont {Landgren}},\ }\href@noop
  {} {\bibfield  {journal} {\bibinfo  {journal} {J. Vac. sci. Technol. B}\
  }\textbf {\bibinfo {volume} {19}},\ \bibinfo {pages} {667} (\bibinfo {year}
  {1981})}\BibitemShut {NoStop}%
\bibitem [{\citenamefont {Pilkington}\ and\ \citenamefont
  {Missous}(1999)}]{Pilkington1999}%
  \BibitemOpen
  \bibfield  {author} {\bibinfo {author} {\bibfnamefont {S.}~\bibnamefont
  {Pilkington}}\ and\ \bibinfo {author} {\bibfnamefont {M.}~\bibnamefont
  {Missous}},\ }\href@noop {} {\bibfield  {journal} {\bibinfo  {journal} {J.
  Cryst. Growth}\ }\textbf {\bibinfo {volume} {196}},\ \bibinfo {pages} {1--12}
  (\bibinfo {year} {1999})}\BibitemShut {NoStop}%
\bibitem [{\citenamefont {Yamada}, \citenamefont {Inokawa},\ and\ \citenamefont
  {Takagi}(1984)}]{Yamada1984}%
  \BibitemOpen
  \bibfield  {author} {\bibinfo {author} {\bibfnamefont {I.}~\bibnamefont
  {Yamada}}, \bibinfo {author} {\bibfnamefont {H.}~\bibnamefont {Inokawa}}, \
  and\ \bibinfo {author} {\bibfnamefont {T.}~\bibnamefont {Takagi}},\
  }\href@noop {} {\bibfield  {journal} {\bibinfo  {journal} {J. Appl. Phys.}\
  }\textbf {\bibinfo {volume} {56}},\ \bibinfo {pages} {2746} (\bibinfo {year}
  {1984})}\BibitemShut {NoStop}%
\bibitem [{\citenamefont {McSkimming}\ \emph {et~al.}(2017)\citenamefont
  {McSkimming}, \citenamefont {Alexander}, \citenamefont {Samuels},
  \citenamefont {Arey}, \citenamefont {Arslan},\ and\ \citenamefont
  {Richardson}}]{McSkimming2017}%
  \BibitemOpen
  \bibfield  {author} {\bibinfo {author} {\bibfnamefont {B.}~\bibnamefont
  {McSkimming}}, \bibinfo {author} {\bibfnamefont {A.}~\bibnamefont
  {Alexander}}, \bibinfo {author} {\bibfnamefont {M.}~\bibnamefont {Samuels}},
  \bibinfo {author} {\bibfnamefont {B.}~\bibnamefont {Arey}}, \bibinfo {author}
  {\bibfnamefont {I.}~\bibnamefont {Arslan}}, \ and\ \bibinfo {author}
  {\bibfnamefont {C.}~\bibnamefont {Richardson}},\ }\href@noop {} {\bibfield
  {journal} {\bibinfo  {journal} {J. Vac. Sci. Technol. A}\ }\textbf {\bibinfo
  {volume} {35}},\ \bibinfo {pages} {2} (\bibinfo {year} {2017})}\BibitemShut
  {NoStop}%
\bibitem [{\citenamefont {Yan}\ \emph {et~al.}(2018)\citenamefont {Yan},
  \citenamefont {Khalsa}, \citenamefont {Vishwanath}, \citenamefont {Han},
  \citenamefont {Wright}, \citenamefont {Rouvimov}, \citenamefont {Katzer},
  \citenamefont {Nepal}, \citenamefont {Downey}, \citenamefont {Muller},
  \citenamefont {Xing}, \citenamefont {Meyer},\ and\ \citenamefont
  {Jena}}]{Yan2018}%
  \BibitemOpen
  \bibfield  {author} {\bibinfo {author} {\bibfnamefont {R.}~\bibnamefont
  {Yan}}, \bibinfo {author} {\bibfnamefont {G.}~\bibnamefont {Khalsa}},
  \bibinfo {author} {\bibfnamefont {S.}~\bibnamefont {Vishwanath}}, \bibinfo
  {author} {\bibfnamefont {Y.}~\bibnamefont {Han}}, \bibinfo {author}
  {\bibfnamefont {J.}~\bibnamefont {Wright}}, \bibinfo {author} {\bibfnamefont
  {S.}~\bibnamefont {Rouvimov}}, \bibinfo {author} {\bibfnamefont {D.~S.}\
  \bibnamefont {Katzer}}, \bibinfo {author} {\bibfnamefont {N.}~\bibnamefont
  {Nepal}}, \bibinfo {author} {\bibfnamefont {B.}~\bibnamefont {Downey}},
  \bibinfo {author} {\bibfnamefont {D.}~\bibnamefont {Muller}}, \bibinfo
  {author} {\bibfnamefont {H.}~\bibnamefont {Xing}}, \bibinfo {author}
  {\bibfnamefont {D.}~\bibnamefont {Meyer}}, \ and\ \bibinfo {author}
  {\bibfnamefont {D.}~\bibnamefont {Jena}},\ }\href@noop {} {\bibfield
  {journal} {\bibinfo  {journal} {Nature}\ }\textbf {\bibinfo {volume} {555}},\
  \bibinfo {pages} {183--189} (\bibinfo {year} {2018})}\BibitemShut {NoStop}%
\bibitem [{\citenamefont {Smith}, \citenamefont {Kroger},\ and\ \citenamefont
  {Jillie}(1983)}]{Smith1983}%
  \BibitemOpen
  \bibfield  {author} {\bibinfo {author} {\bibfnamefont {L.}~\bibnamefont
  {Smith}}, \bibinfo {author} {\bibfnamefont {H.}~\bibnamefont {Kroger}}, \
  and\ \bibinfo {author} {\bibfnamefont {D.}~\bibnamefont {Jillie}},\
  }\href@noop {} {\bibfield  {journal} {\bibinfo  {journal} {IEEE T. Magn.}\
  }\textbf {\bibinfo {volume} {19}},\ \bibinfo {pages} {787--790} (\bibinfo
  {year} {1983})}\BibitemShut {NoStop}%
\bibitem [{\citenamefont {Olaya}\ \emph {et~al.}(2009)\citenamefont {Olaya},
  \citenamefont {Dresselhaus}, \citenamefont {Benz}, \citenamefont
  {Bjarnason},\ and\ \citenamefont {Grossman}}]{Olaya2009}%
  \BibitemOpen
  \bibfield  {author} {\bibinfo {author} {\bibfnamefont {D.}~\bibnamefont
  {Olaya}}, \bibinfo {author} {\bibfnamefont {P.}~\bibnamefont {Dresselhaus}},
  \bibinfo {author} {\bibfnamefont {S.}~\bibnamefont {Benz}}, \bibinfo {author}
  {\bibfnamefont {J.}~\bibnamefont {Bjarnason}}, \ and\ \bibinfo {author}
  {\bibfnamefont {E.}~\bibnamefont {Grossman}},\ }\href@noop {} {\bibfield
  {journal} {\bibinfo  {journal} {IEEE Trans. Appl. Supercond.}\ }\textbf
  {\bibinfo {volume} {19}},\ \bibinfo {pages} {144--148} (\bibinfo {year}
  {2009})}\BibitemShut {NoStop}%
\bibitem [{\citenamefont {Gibson}\ and\ \citenamefont
  {Meservey}(1985)}]{Gibson1985}%
  \BibitemOpen
  \bibfield  {author} {\bibinfo {author} {\bibfnamefont {G.}~\bibnamefont
  {Gibson}}\ and\ \bibinfo {author} {\bibfnamefont {R.}~\bibnamefont
  {Meservey}},\ }\href@noop {} {\bibfield  {journal} {\bibinfo  {journal} {J.
  Appl. Phys.}\ }\textbf {\bibinfo {volume} {58}},\ \bibinfo {pages} {1584}
  (\bibinfo {year} {1985})}\BibitemShut {NoStop}%
\bibitem [{\citenamefont {Huffelen}\ \emph {et~al.}(1993)\citenamefont
  {Huffelen}, \citenamefont {Klapwijk}, \citenamefont {Heslinga}, \citenamefont
  {de~Boor},\ and\ \citenamefont {van~der Post}}]{Huffelen1993}%
  \BibitemOpen
  \bibfield  {author} {\bibinfo {author} {\bibfnamefont {W.}~\bibnamefont
  {Huffelen}}, \bibinfo {author} {\bibfnamefont {T.}~\bibnamefont {Klapwijk}},
  \bibinfo {author} {\bibfnamefont {D.}~\bibnamefont {Heslinga}}, \bibinfo
  {author} {\bibfnamefont {M.}~\bibnamefont {de~Boor}}, \ and\ \bibinfo
  {author} {\bibfnamefont {N.}~\bibnamefont {van~der Post}},\ }\href@noop {}
  {\bibfield  {journal} {\bibinfo  {journal} {Phys. Rev. B}\ }\textbf {\bibinfo
  {volume} {47}},\ \bibinfo {pages} {9} (\bibinfo {year} {1993})}\BibitemShut
  {NoStop}%
\bibitem [{\citenamefont {Kreuzer}, \citenamefont {Wegscheider},\ and\
  \citenamefont {Weiss}(2001)}]{Kreuzer2001}%
  \BibitemOpen
  \bibfield  {author} {\bibinfo {author} {\bibfnamefont {S.}~\bibnamefont
  {Kreuzer}}, \bibinfo {author} {\bibfnamefont {W.}~\bibnamefont
  {Wegscheider}}, \ and\ \bibinfo {author} {\bibfnamefont {D.}~\bibnamefont
  {Weiss}},\ }\href@noop {} {\bibfield  {journal} {\bibinfo  {journal} {J.
  Appl. Phys}\ }\textbf {\bibinfo {volume} {89}},\ \bibinfo {pages} {6751}
  (\bibinfo {year} {2001})}\BibitemShut {NoStop}%
\bibitem [{\citenamefont {Moser}\ \emph {et~al.}(2002)\citenamefont {Moser},
  \citenamefont {Wegscheider}, \citenamefont {Weiss}, \citenamefont {Bichler},\
  and\ \citenamefont {Schuh}}]{Kreuzer2002}%
  \BibitemOpen
  \bibfield  {author} {\bibinfo {author} {\bibfnamefont {S.~K.~J.}\
  \bibnamefont {Moser}}, \bibinfo {author} {\bibfnamefont {W.}~\bibnamefont
  {Wegscheider}}, \bibinfo {author} {\bibfnamefont {D.}~\bibnamefont {Weiss}},
  \bibinfo {author} {\bibfnamefont {M.}~\bibnamefont {Bichler}}, \ and\
  \bibinfo {author} {\bibfnamefont {D.}~\bibnamefont {Schuh}},\ }\href@noop {}
  {\bibfield  {journal} {\bibinfo  {journal} {Appl. Phys. Lett.}\ }\textbf
  {\bibinfo {volume} {80}},\ \bibinfo {pages} {4582} (\bibinfo {year}
  {2002})}\BibitemShut {NoStop}%
\bibitem [{\citenamefont {Zenger}\ \emph {et~al.}(2004)\citenamefont {Zenger},
  \citenamefont {Moser}, \citenamefont {Kreuzer}, \citenamefont {Wegscheider},\
  and\ \citenamefont {Weiss}}]{Zenger2004}%
  \BibitemOpen
  \bibfield  {author} {\bibinfo {author} {\bibfnamefont {M.}~\bibnamefont
  {Zenger}}, \bibinfo {author} {\bibfnamefont {J.}~\bibnamefont {Moser}},
  \bibinfo {author} {\bibfnamefont {S.}~\bibnamefont {Kreuzer}}, \bibinfo
  {author} {\bibfnamefont {W.}~\bibnamefont {Wegscheider}}, \ and\ \bibinfo
  {author} {\bibfnamefont {D.}~\bibnamefont {Weiss}},\ }\href@noop {}
  {\bibfield  {journal} {\bibinfo  {journal} {J. Phys. Condens. Matter}\
  }\textbf {\bibinfo {volume} {16}},\ \bibinfo {pages} {48} (\bibinfo {year}
  {2004})}\BibitemShut {NoStop}%
\bibitem [{\citenamefont {Cho}(1976)}]{Cho1976}%
  \BibitemOpen
  \bibfield  {author} {\bibinfo {author} {\bibfnamefont {A.}~\bibnamefont
  {Cho}},\ }\href@noop {} {\bibfield  {journal} {\bibinfo  {journal} {J. Appl.
  Phys}\ }\textbf {\bibinfo {volume} {47}},\ \bibinfo {pages} {7} (\bibinfo
  {year} {1976})}\BibitemShut {NoStop}%
\bibitem [{\citenamefont {Ohtake}\ \emph {et~al.}(2004)\citenamefont {Ohtake},
  \citenamefont {Kocan}, \citenamefont {Seino}, \citenamefont {Schmidt},\ and\
  \citenamefont {Koguchi}}]{Ohtake2004}%
  \BibitemOpen
  \bibfield  {author} {\bibinfo {author} {\bibfnamefont {A.}~\bibnamefont
  {Ohtake}}, \bibinfo {author} {\bibfnamefont {P.}~\bibnamefont {Kocan}},
  \bibinfo {author} {\bibfnamefont {K.}~\bibnamefont {Seino}}, \bibinfo
  {author} {\bibfnamefont {W.}~\bibnamefont {Schmidt}}, \ and\ \bibinfo
  {author} {\bibfnamefont {N.}~\bibnamefont {Koguchi}},\ }\href@noop {}
  {\bibfield  {journal} {\bibinfo  {journal} {Phys. Rev. Lett.}\ }\textbf
  {\bibinfo {volume} {93}},\ \bibinfo {pages} {266101} (\bibinfo {year}
  {2004})}\BibitemShut {NoStop}%
\bibitem [{\citenamefont {Pasquariello}\ and\ \citenamefont
  {Hjart}(2002)}]{Pasquariello2002}%
  \BibitemOpen
  \bibfield  {author} {\bibinfo {author} {\bibfnamefont {D.}~\bibnamefont
  {Pasquariello}}\ and\ \bibinfo {author} {\bibfnamefont {K.}~\bibnamefont
  {Hjart}},\ }\href@noop {} {\bibfield  {journal} {\bibinfo  {journal} {IEEE J.
  Sel. Top. Quant.}\ }\textbf {\bibinfo {volume} {8}},\ \bibinfo {pages} {1}
  (\bibinfo {year} {2002})}\BibitemShut {NoStop}%
\bibitem [{\citenamefont {Collins}(1997)}]{Collins1997}%
  \BibitemOpen
  \bibfield  {author} {\bibinfo {author} {\bibfnamefont {S.}~\bibnamefont
  {Collins}},\ }\href@noop {} {\bibfield  {journal} {\bibinfo  {journal} {J.
  Electrochem. Soc.}\ }\textbf {\bibinfo {volume} {144}},\ \bibinfo {pages}
  {2242--2262} (\bibinfo {year} {1997})}\BibitemShut {NoStop}%
\bibitem [{\citenamefont {Kim}, \citenamefont {Lim},\ and\ \citenamefont
  {Yang}(1998)}]{Kim1998}%
  \BibitemOpen
  \bibfield  {author} {\bibinfo {author} {\bibfnamefont {J.}~\bibnamefont
  {Kim}}, \bibinfo {author} {\bibfnamefont {D.}~\bibnamefont {Lim}}, \ and\
  \bibinfo {author} {\bibfnamefont {G.}~\bibnamefont {Yang}},\ }\href@noop {}
  {\bibfield  {journal} {\bibinfo  {journal} {J. Vac. Sci. Technol. B}\
  }\textbf {\bibinfo {volume} {16}},\ \bibinfo {pages} {558--560} (\bibinfo
  {year} {1998})}\BibitemShut {NoStop}%
\bibitem [{\citenamefont {Scigliuzzo}\ \emph {et~al.}(2020)\citenamefont
  {Scigliuzzo}, \citenamefont {Bruhat}, \citenamefont {Bengtsson},
  \citenamefont {Burnett}, \citenamefont {Roudsari},\ and\ \citenamefont
  {Delsing}}]{Scigliuzzo2020}%
  \BibitemOpen
  \bibfield  {author} {\bibinfo {author} {\bibfnamefont {M.}~\bibnamefont
  {Scigliuzzo}}, \bibinfo {author} {\bibfnamefont {L.}~\bibnamefont {Bruhat}},
  \bibinfo {author} {\bibfnamefont {A.}~\bibnamefont {Bengtsson}}, \bibinfo
  {author} {\bibfnamefont {J.}~\bibnamefont {Burnett}}, \bibinfo {author}
  {\bibfnamefont {A.}~\bibnamefont {Roudsari}}, \ and\ \bibinfo {author}
  {\bibfnamefont {P.}~\bibnamefont {Delsing}},\ }\href@noop {} {\bibfield
  {journal} {\bibinfo  {journal} {New J. Phys.}\ }\textbf {\bibinfo {volume}
  {22}},\ \bibinfo {pages} {053027} (\bibinfo {year} {2020})}\BibitemShut
  {NoStop}%
\end{thebibliography}%

\end{document}